\documentclass[prd,aps,12pt,amssymb,preprintnumbers,amsmath,amsfonts,nofootinbib]{revtex4-2}
\usepackage{amsmath}

\newcommand{\ba}{\begin{eqnarray}}
\newcommand{\ea}{\end{eqnarray}}

\newcommand{\be}{\begin{equation}}
\newcommand{\ee}{\end{equation}}
\newcommand{\pa}{\partial}

\newcommand{\nn}{\nonumber}

\newcommand{\Eq}[1]{Eq.~\eqref{#1}}

\newcommand{\OO}{{\cal {O}}}
\newcommand{\q}{{\bf {q}}}

\newcommand{\itq}{{\int_{\q} \, }}
\newcommand{\ifq}{{\int_Q} \, }
             
\usepackage{graphicx}
\usepackage{slashed}

\usepackage{hyperref}  

\begin{document}
                                                                                
\date{\today}

\title{Power corrections and gradient expansion in QED transport theory} 
\author{Stefano Carignano}
\email{carignano@ice.cat}
\affiliation{Instituto de Ciencias del Espacio (ICE, CSIC) \\
C. Can Magrans s.n., 08193 Cerdanyola del Vall\`es, Catalonia, Spain
and \\
 Institut d'Estudis Espacials de Catalunya (IEEC) \\
 C. Gran Capit\`a 2-4, Ed. Nexus, 08034 Barcelona, Spain
}
\author{Cristina Manuel}
\email{cmanuel@ice.csic.es}
\affiliation{Instituto de Ciencias del Espacio (ICE, CSIC) \\
C. Can Magrans s.n., 08193 Cerdanyola del Vall\`es, Catalonia, Spain
and \\
 Institut d'Estudis Espacials de Catalunya (IEEC) \\
 C. Gran Capit\`a 2-4, Ed. Nexus, 08034 Barcelona, Spain
}

\begin{abstract}
 
The  hard thermal loop (HTL) effective field theory of QED can be derived from the classical limit of transport theory, corresponding to the leading term in a gradient expansion of the quantum approach.
In this paper, we show that power corrections to the HTL effective Lagrangian of QED can also be obtained 
from transport theory by including higher orders in such gradient expansion.
The gradient expansion is increasingly infrared (IR) divergent, but the correction that we compute is IR finite.
We employ dimensional regularization, and show that this result comes after a cancellation of divergencies 
between the vacuum and medium contributions. While the transport framework is an effective field theory of the long distance physics of the plasma, we show that it correctly reproduces the correct QED ultraviolet divergencies associated with the photon wave function renormalization.

\end{abstract}

\maketitle

\section{Introduction}

The characterization of relativistic hot plasmas is an extensive field of investigation which has applications ranging
from nuclear physics to condensed matter and cosmology.
{Within the weak-coupling regime,  the presence of a large scale given by the temperature $T$ and a small gauge coupling constant $e$
allows to separate between ``hard" ($\sim T$) and ``soft" $(\sim e T)$ scales. In this regime, perturbative} 
thermal field theory is applicable
and allows for well-defined computations of different physical quantities in powers of the gauge coupling constant.
Unfortunately, this  requires the resummations of  a set of Feynman diagrams, the so called hard thermal loops (HTL)  \cite{Pisarski:1988vd,Braaten:1989mz},  which makes 
the perturbative approach not so simple to apply.  The resummed HTL effective theory is used to compute the soft contribution to every physical observable (see eg. \cite{Laine:2016hma,Ghiglieri:2020dpq} for recent reviews), while the bare theory is used to evaluate the hard contribution.

Recently, a systematic QED calculation has extended the effective HTL Lagrangian  by including next-to-leading order corrections.  In the photon sector there are two contributions required for this calculation: One originates 
from two-loop diagrams \cite{Carignano:2019ofj}, while the other from power corrections to the one-loop HTL result   \cite{Manuel:2016wqs,Carignano:2017ovz}.  While the first are standard perturbative corrections 
proportional to $e^2$, the second provide corrections of order $(\ell/T)^2$ to the HTL result, $\ell$ being the momentum of the photon: for soft momenta, they are thus $ e^2 $ corrections and thus comparable with the two-loop contributions, although they can also become the leading correction 
 if $\ell$ lies in some intermediate regime between soft and hard scales.  
Actually, the power corrections were first computed from
the on-shell effective theory, which takes advantage of the hierarchy of scales of the hard and soft sectors
\cite{Manuel:2016cit,Manuel:2014dza}.
For the fermionic sector, the power corrections have been also computed \cite{Carignano:2017ovz}, while the two-loop diagrams have not been yet evaluated.
 
 A completely different approach to study relativistic plasmas is based on using  transport theory \cite{Elze:1989un,Blaizot:2001nr,Litim:2001db}.
It is actually possible to show that the physics of the soft scales can be studied within this framework, reproducing the HTL Feynman diagrams 
\cite{Blaizot:1992gn,Blaizot:1993zk,Blaizot:1993be,Kelly:1994ig,Kelly:1994dh}.
This {effective} approach has been used successfully to study numerically {the  dynamical evolution}  of relativistic plasmas, as lattice techniques are in principle only amenable to study their thermodynamics.

Over the past few years, a renewed interest in the role of quantum effects in transport theory has been motivated by the study of chiral plasmas, characterized 
by an imbalance between different species of massless fermions. 
In this particular context, quantum corrections 
can be incorporated in the kinetic equations
and generate several novel effects, such as the chiral magnetic and vortical effects (see eg. \cite{Stephanov:2012ki,Son:2012zy,Gao:2012ix,Hidaka:2016yjf,Mueller:2017arw,Huang:2018wdl,Carignano:2018gqt,Lin:2019ytz,Weickgenannt:2019dks}). 
We will not consider the situation of
chirally imbalanced systems in this manuscript, though.

Instead, 
 in this work, we aim at showing that the power corrections to the HTL Lagrangian of QED can also be obtained  via the transport approach.
While the HTL physics is recovered from the leading term in a gradient expansion of the transport approach, which describes the
classical limit of the theory, the power corrections are obtained by keeping the first non-vanishing correction in a gradient expansion.
 Focusing on the photon sector,  we will discuss how the connection between the gradient expansion and power corrections naturally emerges.
 While we will be working in natural units
and discuss the different approximations in terms of ratios of different energy scales in the system, it is also possible to see that the relevant terms we will use correspond to $\hbar^2$ corrections in the transport approach.

This paper is organized as follows: in Sec.~\ref{sec:transport} we review the derivation of the transport and constraint equations including higher order terms of the gradient expansion. In Sec.~\ref{sec:powKT}
we then show how the power corrections can be derived from the transport equation. We finally discuss our results and outlooks in Sec.~\ref{sec:conclusions}. Some technical details on 
the regularization of possible divergencies using dimensional regularization are discussed in the Appendix.
We use natural units  $\hbar=c= k_B=1$ and  the  metric   $g^{\mu\nu} = diag(1,-1,-1,-1)$.

\section{Transport and constraint equations for QED  }
\label{sec:transport}

As a first step, let us briefly rederive the transport and constraint equations which will be used to recover the power corrections. For more in-depth discussions, we refer to the vast literature on the subject,
see eg. \cite{Vasak:1987um,Blaizot:1992gn,Zhuang:1995pd} or \cite{Gao:2020pfu} for a recent review.

{We consider a relativistic QED plasma at high temperature $T$, and thus
assume that the fermions can be treated as massless. }
Our focus will be on the fermionic Wigner function $W(X,q)$, which amounts to the quantum analogue of  the particle distribution function in transport theory, and is defined as the Fourier transform of the ensemble average
of the two-point correlator
\be
\label{wigner}
 W(X,q)= \int\frac{d^4 s}{(2\pi)^4}
e^{-iq\cdot s} W(x,y) 
=
 \int\frac{d^4 s}{(2\pi)^4}
e^{-iq\cdot s}
\left\langle \bar\psi(X+\frac{s}{2})U(X+\frac{s}{2},X-\frac{s}{2})
\psi(X-\frac{s}{2})\right\rangle\,,
\ee
where $\psi$ is the fermionic field, $X=(x+y)/2$ and $s=x-y$ are the center-of-mass and relative coordinates, respectively, and $U$ is the link operator ensuring the gauge invariance of the Wigner function,
\be
\label{link}
U\left(X+\frac{s}{2},X-\frac{s}{2}\right) = 
 \exp\left\lbrace {-ie s^\mu\int_0^1 dz \, A_\mu\left(X-\frac{s}{2}+zs\right)}\right\rbrace \,,
\ee
where the path connecting $x$ and $y$ is taken to be a straight line. 
Since  in the following we will consider self-consistent mean gauge fields  \cite{Blaizot:1992gn} in order to reconstruct the HTLs and their power corrections, we need not care about subtleties associated with 
the path ordering~\cite{Vasak:1987um}.

Note that we do not include normal ordering in the definition of the Wigner function, as vacuum contributions will play an important role in our derivation of the power corrections.

We now take the Dirac equation for massless fermions and its adjoint acting on $W(x,y)$,
\be
\slashed{D}_x W(x,y) = W(x,y)  \slashed{D}_y^\dagger = 0  \,, \qquad \; D_x^\mu = \pa_x^\mu + i e A^\mu(x) \,, 
\label{eq:eqD}
\ee 
 and after squaring them, 
we can build a sum and a difference equations, which will give us a constraint equation, enforcing the 
on-shell condition for the quasiparticles, and the transport equation, respectively \cite{Vasak:1987um,Blaizot:1992gn,Zhuang:1995pd}:
\begin{align}
\label{initial-eqs}
(D_x^2 \mp D_y^{*2}) W(x,y) - \frac{e}{2} \Big[ F_{\mu\nu}(x) \sigma^{\mu\nu} W(x,y) \mp F_{\mu\nu}(y)  W(x,y) \sigma^{\mu\nu} \Big] = 0 \,,
\end{align}
with $F_{\mu\nu} = \pa_\mu A_\nu - \pa_\nu A_\mu $ the electromagnetic stress tensor.

\subsection{The transport equation}

Focusing first on the difference equation associated with Eq.~(\ref{initial-eqs}), 
we derive the transport equation for a fermion in the plasma, recalling that in order to have a well-defined quasiparticle interpretation, 
its momentum $q$ must be hard. We perform
a Wigner transformation and a gradient expansion around $X$. 
In principle, this can be worked to all orders \cite{Vasak:1987um}, although for our purposes we will truncate the expansion
at order ${\cal O}(\pa_X^3)$.
 We obtain 
 (in the following, for brevity, we will omit the $X$ on the corresponding derivatives, ie. $\pa \equiv \pa_X$) 
  \begin{align}
 & (q\cdot \pa - e q\cdot F\cdot \pa_q)W(X,q) + \frac{i e}{4}F_{\mu\nu} [\sigma^{\mu\nu}, W] \; - \frac{e}{8} \Delta F_{\mu\nu}(X) \big\lbrace \sigma^{\mu\nu}, W \big\rbrace \nn\\
 & = -\Big[ \frac{e}{24} q^\mu (\Delta^2 F_{\mu\nu}) \pa_q^\nu + \frac{e}{12}(\Delta F^{\lambda\nu}) \pa^q_\lambda (\pa_\nu - e F_{\nu\beta}\pa_q^\beta) \Big] W
 +  \frac{i e}{32} \Delta^2 F_{\mu\nu} [\sigma^{\mu\nu}, W]  \,,
 \label{eq:transporth2}
 \end{align}
where  {$\Delta= \pa \cdot \pa_q$,  with $\pa$ only acting on $F^{\mu \nu}$}. 
 In order for the expansion to be well behaved,
we thus demand that the  pieces involving the $\Delta$ operator be small. In Fourier modes, if
$\ell$ is the momenta associated with the electromagnetic field,  this implies
\be
\Delta \ll 1 \ , \qquad  \frac{\ell}{q} \ll 1 \,.  
\ee
At this point we can also recall that the gradient expansion can also be seen as a quantum expansion in $\hbar$: in this sense, the anti-commutator piece in \Eq{eq:transporth2} is of order $\hbar$, 
 whereas all the  terms {on its right-hand side} are of order $\hbar^2$ \cite{Vasak:1987um}. These latter ones will be the pieces required to obtain the power corrections, as we will show in the following.

 Upon inspecting the transport equation, recalling that the Wigner function is a matrix in Dirac space, it is clear that its components can get in principle mixed. 
In order to continue, we can perform a decomposition of its Dirac structure and
find individual equations for its components. More specifically, we write \cite{Vasak:1987um}
 \be
W = {\cal F} + i \gamma^5 {\cal P} + \gamma^\alpha V_\alpha + \gamma^5 \gamma^\alpha A_\alpha + \frac{1}{2} \sigma^{\mu\nu} S_{\mu\nu} \, ,
\label{eq:WFdecomp}
\ee
where ${\cal F}, {\cal P}, V_\alpha, A_\alpha$ and  $S_{\mu \nu}$ stand for the scalar, pseudoscalar, vector, axial-vector and tensorial components
of the Wigner function.

Of particular interest for us in the derivation of the hard thermal loops and their power corrections 
will be the vectorial component $V^\alpha$, so if we now trace our kinetic equation multiplied by a $\gamma^\alpha$,
pieces proportional to the anticommutator $\{ \sigma^{\mu\nu} , W \} $ drop due to symmetry and 
we  get to the following equation\footnote{Note that, compared to  Eq. (4.28) in \cite{Vasak:1987um}, this equation contains extra $\Delta^2$ pieces which are required to get the full $\hbar^2$  or power corrections.}
\be
\Bigg\lbrace \Big[  q\cdot \pa - e q\cdot F\cdot \pa_q \; {{{{+}}}} \; \frac{e}{24} q^\mu (\Delta^2 F_{\mu\nu}) \pa_q^\nu + \frac{e}{12}(\Delta F^{\lambda\nu}) \pa^q_\lambda  (\pa_\nu - e F_{\nu\sigma}\pa_q^\sigma) \Big] g^{\alpha\beta} - e \Big(1- \frac{1}{8}\Delta^2\Big) F^{\alpha\beta} \Bigg\rbrace V_\beta = 0 \,.
\label{eq:KEV}
\ee
Unlike for the case of chirally imbalanced plasmas, we can thus see that the first non-vanishing corrections to \Eq{eq:KEV} are  $\OO(\hbar^2)$, as
$\OO(\hbar)$ corrections cancel out when summing over the fermion spin, and the axial components of the Wigner function do not play a role in our specific computation.

\subsection{The dispersion relation}
If we now consider the sum equation in Eq.~(\ref{initial-eqs})  we can 
 we can get a constraint obeyed by the Wigner function, from which one can obtain the fermionic dispersion relation.
After Wigner transforming, and truncating to order $\pa_X^3$, this becomes 
\begin{align}
\Big[ & Q^2 - \frac{e}{6} q^\mu \Delta F_{\mu\nu} \pa_q^\nu -  \frac{e}{12} \pa^\mu F_{\mu\nu} \pa_q^\nu - \frac{1}{4} \Big(\pa_\mu - e F_{\mu\nu} \pa_q^\nu)^2 \Big] W(X,q) \nn\\
& = \frac{e}{4} F_{\mu\nu} \{ \sigma^{\mu\nu} ,W\} - i \frac{e}{8} \Delta F_{\mu\nu} [\sigma^{\mu\nu}, W ]  \, ,
\label{eq:disprel}
\end{align}
where $Q^2 = q_\mu q^\mu$, 
and after projecting on the vector component, we obtain
\begin{align}
\Big[ & Q^2 - \frac{e}{6} q^\mu \Delta F_{\mu\nu} \pa_q^\nu -  \frac{e}{12} \pa^\mu F_{\mu\nu} \pa_q^\nu - \frac{1}{4} \big(\pa_\mu - e F_{\mu\nu} \pa_q^\nu\big)^2 \Big] V^\alpha + \frac{e}{2} \Delta F^{\alpha\beta}V_\beta = 0 \,,
\label{eq:disprelV}
\end{align}
which provides a correction to the simple on-shell relation for massless particles $Q^2 = 0$. As we will argue in the following, however, this does not play a role 
in the derivation of the power corrections.

 \section{Power corrections to the photon HTL from kinetic theory} 
 \label{sec:powKT}
In this section we show how,
by making use of the transport equation obtained in the previous section, we can derive the power corrections to the HTL Lagrangian. 
For this, we start from the electromagnetic current associated with the fermionic field, 
 \be
 j^\mu = 4 e \int \frac{d^4q }{(2\pi)^4} V^\mu \,,
 \ee
 where $V^\mu$ again is the vector component of the fermionic Wigner function.

 We now split $V^\mu$ into 
  \be
 V^\mu = V^\mu_{(0)} +  e \Big(  V^\mu_{{\rm HTL}} + V^\mu_{{\rm pow}} + \dots \Big) + e^2 (  \dots )  \,,
 \label{eq:Vsplit}
 \ee
 where, aside from the leading-order term $V_{(0)}$ we singled out the leading $\OO(e)$ contribution, describing  the well-known HTL result, 
 and the first non-vanishing gradient correction to it, which will amount to the power corrections to the HTL, as we will show in what follows. 
In terms of explicit factors of the gauge coupling, this is still an $\OO(e)$ term,
 although in the realm of validity of the gradient expansion, it will be suppressed by  $\ell/q \ll 1$ factors, where $\ell$ is the soft photon momentum and $q\sim T$.  If we interpret the gradient expansion as an expansion in terms of $\hbar$  \cite{Vasak:1987um,Zhuang:1995pd}, the power corrections 
 can be seen as an $\hbar^2$ correction to the HTL result.

Since in the computation of the electromagnetic current
 when going to higher orders in the gradient expansion we  meet with diverging contributions, in the following we will employ dimensional regularization (DR) 
  and perform all our integrals in $D = d+1 = 4+2\epsilon$ dimensions.

As first step, let us rederive the HTL current.
In thermal equilibrium we have { (see, for example \cite{Blaizot:1992gn})}
\be
 V_{(0)}^\alpha = q^\alpha \, 2\pi \, sgn(q^0) \, \delta(Q^2) f_F(q^0) \equiv q^\alpha G_0(q) \ ,
\ee
 where $f_F = [1+ \exp(q^0/T)]^{-1}$ is the  fermionic occupation number, so we get to 
\be
  q\cdot \pa  V^\alpha_{\rm HTL}  =  e \Big[ g^{\alpha}_{\;\; \beta} q\cdot F\cdot \pa_q  + F^{\alpha}_{\;\; \beta} \Big] V_{(0)}^\beta  \,,
  \label{eq:KEhtl}
\ee
where we used that we are near equilibrium and $V^{(0)}$ is space-independent.  We can now solve for $V_{\rm HTL} $, and plugging our result in the current, 
after integrating by parts in order to avoid having 
 derivatives acting on $G_0$, we get 
  \be
 j^\mu_{{\rm HTL} }  = - 8\pi e^2  \ifq \frac{1}{2q} \Big[ \delta(q^0 - q) - \delta(q^0+q) \Big] f_F(q^0) \Big[ \frac{q_\lambda }{q\cdot \pa} F^{\lambda\mu} -  \pa_\nu \frac{q^\mu q_\lambda} {(q\cdot \pa)^2}  F^{\lambda\nu} \Big]  \,,
 \label{eq:intermHTL2}
 \ee
 with $q = \vert{\bf q}\vert$.  Note that 
   when changing the dimensions of space-time, in order to keep the gauge coupling dimensionless we multiply it by a factor 
$ \nu^{3 -d}$, where  $\nu$ is a renormalization scale. For brevity, we introduced the shorthand notation
\be
\int_Q \equiv \int \frac{dq^0}{2\pi} \int_\q \equiv   \nu^{3 -d} \int \frac{dq^0}{2\pi} \int \frac{d^dq }{(2\pi)^d} .
 \ee

 After employing the delta functions to perform the $q^0$ integration, we recover the particle and antiparticle contributions, $q^\mu = q v^\mu = q(1,{\bf v}) $  and $q^\mu = -q \tilde{v}^\mu = -q(1,-{\bf v})$, ${\bf v}^2 = 1$.
 Making use of the symmetries in the angular integration we now send $\tilde{v}^\mu \to v^\mu$, and using 
 that $f_F(-q) = 1-f_F(q)$ we get 
   \be
 j^\mu_{{\rm HTL} }  = 2 e^2  \itq \frac{1 - 2 f_F(q)}{q}  \Big[ \frac{v_\lambda }{v\cdot \pa} F^{\lambda\mu} -  \pa_\nu \frac{v^\mu v_\lambda} {(v\cdot \pa)^2}  F^{\lambda\nu} \Big]  \,,
 \label{eq:jHTL}
 \ee
 and if we now move to momentum space and compute the photon polarization tensor as
 \be
 \Pi^{\mu\nu}_{{\rm HTL} } (\ell) = \frac{\delta j^{\mu}_{\rm HTL}}{\delta A_\nu} =  2 e^2  \itq \frac{1 - 2 f_F(q)}{q} \Big[ g^{\mu\nu} - \frac{v^\mu \ell^\nu +v^\nu \ell^\mu }{v\cdot \ell} 
 + \frac{v^\mu v^\nu L^2}{(v\cdot \ell)^2} \Big] \, , 
 \label{eq:piHTL}
 \ee
 with $L^2 = \ell^\mu\ell_\mu$, we recover the familiar HTL expression, see for example \cite{Pisarski:1997cp}. 
 {Note that the expression above contains  in principle both  thermal and vacuum contributions, the latter however do not contribute here, as DR sets  
  scale-less integrals to zero. }

 We are now ready to compute the power corrections to the HTL polarization tensor. For this, we start from \Eq{eq:KEV} truncated at ${\cal{O}}(e)$ and again split the vector component of the Wigner function as 
in \Eq{eq:Vsplit}.  Using the fact that 
  $V^\mu_{HTL}$ satisfies \Eq{eq:KEhtl} to simplify the equation, we are left with 
\be
q\cdot \pa V_{\rm{pow}}^\alpha + e\Big[
\frac{1}{24} q^\mu (\Delta^2 F_{\mu\nu}) \pa_q^\nu 
 g^{\alpha\beta}+ \frac{1}{8}\Delta^2 F^{\alpha\beta}\Big] V^{(0)}_\beta = 0 \,,
\label{eq:KEpow}
\ee
where we used that we are near equilibrium and that $V^{(0)}$ is space-independent. Further, we have dropped pieces of order $e^2$ which would
only be needed if we were to reproduce three-point functions.

 We can already note at this point that the terms appearing in \Eq{eq:KEpow}
 are quadratic in the operator $\Delta$, which will translate in quadratic  corrections in powers of $\ell/q$. 
  This is a consequence of the symmetries of the system, in particular parity $P$ and $CP$, where $C$ is charge conjugation, which enforce that no odd powers of $\ell$ can occur.

Now we can use this result to extract 
 \begin{align}
 j^\mu_{\rm{pow}} &  = 4 e \ifq V^\mu_{{\rm pow}} = 4 e^2  \ifq \Big( \frac{ \pa^4}{4(q\cdot \pa)^2} \Big)  \Big[ \frac{q_\lambda }{q\cdot \pa} F^{\lambda\mu} -  \pa_\nu \frac{q^\mu q_\lambda} {(q\cdot \pa)^2}  F^{\lambda\nu} \Big] G_0(q) \,,
\label{eq:jPOW}
 \end{align}
which is exactly the HTL form of \Eq{eq:intermHTL2} multiplied by the factor $- \frac{ \pa^4}{4(q\cdot \pa)^2}$.

At this point, we should check whether the corrections to the on-shell relation \Eq{eq:disprel} from the gradient expansion play a role at this order in the computation.
With the splitting (\ref{eq:Vsplit}) we then see that the constraint equation 
 for $V_{\rm {pow}}$ is modified by pieces going as $\sim e \Delta G_{(0)}$  stemming from 
the gradient expansion. However, this would only introduce
higher order corrections to the electromagnetic current, which we neglect at the order of the computation we
are carrying out. Thus, we can consider the dispersion law of the free massless case for our computations of the power corrections current.

Employing the leading-order on-shell condition for massless particles,
performing again the $q^0$ integrals and going to momentum space, we thus find the expression for the power corrections to the photon polarization tensor,
 \be
 \Pi^{\mu\nu}_{\rm {pow}}(\ell) = \frac{\delta j^{\mu}_{\rm{pow}}}{\delta A_\nu} =  2 e^2  \itq \frac{1 - 2 f_F(q)}{q^3} \frac{ L^4}{4(v\cdot \ell)^2}  \Big[ g^{\mu\nu} - \frac{v^\mu \ell^\nu +v^\nu \ell^\mu }{v\cdot \ell} + \frac{v^\mu v^\nu L^2}{(v\cdot \ell)^2} \Big] \,,
 \label{eq:piPOW}
 \ee
  and the effective Lagrangian generating it  via $j^\alpha_{\rm pow} = \delta {\cal L}_{\rm pow} / \delta A_\alpha$ 
 is
\be
\label{npc-HTL}
{\cal L}_{\rm pow} = \frac{e^2}{4} \itq  \frac{ 1- 2 f_F(q) }{q^3} \Bigg \{ F_{\rho \alpha}   \frac{v^\alpha v^\beta}
{(v \cdot \partial)^4 } \partial^4 F _{\beta}^{\,\,\rho}  \Bigg \} \,,
\ee
 which is the same result found using diagrammatic methods in \cite{Manuel:2016wqs,Carignano:2017ovz}.

Let us emphasize (see the Appendix) that the momentum integrals of Eqs.~(\ref{eq:piPOW},\ref{npc-HTL})
are infrared finite, as there is a cancellation of the IR singularity associated with the thermal bath with that of the vacuum.
For this reason, it is important to keep the vacuum contribution. The momentum integrals still contain 
a ultraviolet  (UV) divergence, associated with the photon wavefunction renormalization. This is cured with the standard QED 
counterterm 
\be
{\cal L}_{\rm c.t.} = - \frac{ Z(\alpha, \epsilon)}{4} F_{\mu \nu} F^{\mu \nu} 
\ , \qquad
Z  = 1 - \frac{2 \alpha}{ 3 \epsilon \pi} \ ,
\ee
 where $\alpha= e^2/4\pi$ is the electromagnetic fine structure constant.

In Ref.~\cite{Carignano:2017ovz} it was pointed out that the polarization tensor for photon momentum $\ell \ll T$ admits an expansion of the
form 
\be
\left(\frac {L^4}{4 q^2 (v \cdot \ell)^2} \right)^n \ , \qquad n= 1,2, 3 \ldots
\ee
times the integrand of the HTL polarization tensor, Eq.(\ref{eq:piHTL}). Our work suggests that these corrections could also be obtained by
keeping extra terms in the gradient expansion, ie. higher powers of the $\Delta$ operator in the transport approach. Note however that, starting from the next power
correction to the one here computed, the expansion is increasingly  IR divergent. These divergencies are most likely cancelled by computing
{ soft loop contributions to the
polarization tensor}, which require the use of resummed pertubation theory, with modified  vertices and propagators. We will however not address this issue here.

\section{Conclusions}
\label{sec:conclusions}

{While it is well-known that the HTL effective Lagrangian can be derived from the classical limit of transport theory,
in this work we have shown how the power corrections to the HTL can also  be reproduced within this approach, by keeping 
next-to-leading order terms of
the gradient expansion. Although we have worked using natural units,  it is possible to see that
these correspond to pure quantum corrections $\propto \hbar^2$ \cite{Vasak:1987um}.
  In particular, the relevant  new contributions discussed in this work}
are given by $(\pa_X\cdot \pa_q)^2 \sim (\ell/q)^2 $ corrections to the HTL, where $\ell$ is the soft photon momentum and $q \sim T$ a hard momentum.
{No odd terms in the photon momentum can appear in the expansion due to the P and CP symmetries of the plasma we considered here, since we didn't allow for any kind of parity-breaking
mechanism \cite{Nieves:1988qz}.}
This is also reflected in the absence of $\OO(\hbar)$ corrections, which  instead arise when considering chirally imbalanced plasmas.

Let us emphasize again that it is important to keep the vacuum contribution in our framework, as the first gradient expansion correction to the HTL result is only IR finite after including it. 
The vacuum is here treated as another
medium, more specifically, as a Dirac sea of antiparticles. We have seen that with our formulation we can reproduce the correct QED 
photon wavefunction UV divergence. At first, this result might be puzzling, as it is an UV effect, while 
the transport theory is an effective field theory for the long distance physics. However, in the modern formulations of effective field theories, one should not use cutoffs in their definitions, but rather a gauge respectful regulator, such as dimensional regularization. All the momentum scales thus enter in the effective field theory approach.
The scale of the physics that the effective field theory describes results
after performing the different momentum integrals, which are finite if used in dimensional regularization.
Finally, a matching procedure with the full theory  at a given scale should be done. This procedure should be applied to the transport framework as well.

It would be interesting to investigate how higher-order power corrections can be derived from the transport equations, 
as our work suggests that this should be feasible. However, when the photon momenta $\ell$ is soft ($\sim e T$), the power corrections computed here
turn out to be of the same order as a standard perturbative correction that shows up in two-loop  diagrams
\cite{Carignano:2019ofj}. 
{It should also be possible to derive these two-loop corrections of the polarization tensor in transport theory,
although it would likely be much more involved.}

While we have centered our analysis in electromagnetic plasmas, it would be very interesting to generalize our approach to the quark-gluon plasma as well. We hope to report on those efforts in the future.

 \section*{Acknowledgements} 

We have been supported by Ministerio de Ciencia, Investigacion y Universidaddes (Spain) under the project PID2019-110165GB-I00 (MCI/AEI/FEDER, UE),  as well as by the Generalitat de
Catalunya by the project  2017-SGR-929  (Catalonia). This work was also supported by the COST Action CA15213 THOR.

\appendix

\section{DR integrals }
\label{app:DR}

In this appendix we briefly recall some explicit formulas for evaluating our expressions using dimensional regularization. 
We perform 
the momentum integrals $d = 3+2\epsilon$ spatial dimensions, replacing
 \cite{Laine:2016hma}
\be
\int \frac{d^3 q}{(2\pi)^3} \rightarrow 
\int \frac{d^d q}{(2\pi)^d} = \int_0^\infty dq \, q^{d-1} \int \frac{d\Omega_d}{(2\pi)^d} =
\frac{ 4 }{(4\pi)^{\frac{d+1}{2}} \Gamma( \frac{d-1}{2})} \,\int_0^\infty dq \, q^{d-1} \int_{-1}^1 d\cos\theta \sin^{d-3}\theta \ ,
\ee
where $\theta$ parametrizes the angle with respect to an external vector, and $\Gamma(z)$ is the Gamma function.

The relevant radial integral appearing in the power corrections is 
\begin{align}
\nu^{- 2 \epsilon} \int_0^\infty dq q^{-1+2\epsilon} \,f_F (q)  & =
\left(\frac{\nu}{T}\right)^{-2\epsilon} (1-2^{1-2\epsilon})\Gamma (2\epsilon) \zeta (2\epsilon)
=
\frac{1}{4\epsilon} + \frac{1}{2}\ln \left(\frac{\pi Te^{-\gamma_E}}{2 \nu}\right)  + {\cal{O}}(\epsilon) \ ,
\label{DR-fer}
\end{align}
where $\nu$ is the DR scale, $\gamma_E$ is the Euler's constant and $\zeta(z)$ stands for the Riemann zeta function. 

 Note that the above integral in $d=3$ is  IR-divergent, so that regularization is required. However, when adding the vacuum contribution, we find a subtle cancellation of the IR divergences of the vacuum and thermal parts. To see this, one can expand for low momenta $ 1 - 2 f_F(q) \approx  \frac{q}{2T}+ {\cal O}(\frac{q^3}{T^3})$ and thus see that the integral that appear in Eq.~(\ref{eq:jPOW}) is IR finite, but still has the UV divergence associated with the vacuum contribution.

Our expression for the power corrections 
can be written as 
 \be
 \Pi^{\mu\nu}_{\rm {pow}}(\ell) = - e^2 \Big[\frac{1}{\epsilon} + 2 \ln \left(\frac{\pi T e^{-\gamma_E}}{2 \nu}\right)\Big] \frac{L^4}{4}
  \int\frac{d\Omega_d}{(2\pi)^d} \frac{1}{(v\cdot \ell)^2}  \Big[ g^{\mu\nu} - \frac{v^\mu \ell^\nu +v^\nu \ell^\mu }{v\cdot \ell} + \frac{v^\mu v^\nu L^2}{(v\cdot \ell)^2} \Big] \,,
 \label{eq:piPOWDR}
 \ee
and the $1/\epsilon$ pole can be cancelled with the vacuum QED counterterm. This can be seen by evaluating explicitly 
the longitudinal and transverse parts, which give (taking care of keeping the finite pieces arising from the angular integrals in $d$ dimensions \cite{Carignano:2017ovz})

\begin{align}
\Pi^{L}_{{\rm pow}} &= \frac{\alpha}{ 3 \pi }\left[ \frac{l^2}{ \epsilon} +  2 l^2
 \left( \ln \frac{\sqrt{\pi}Te^{- \gamma_E/2}}{2 \nu}  -1 \right) +
\left(2l^2-L^2\right)\left(1- \frac{l_{0}}{2 l}
 \,{\rm ln\,}{\frac{l_0+ l}{l_0-{ l}}} \right) \right] \,, \label{final-L}
\\
 \Pi^{T}_{{\rm pow}}  &=  \frac{ 2\alpha L^2}{3\pi } \left[ \frac{1}{2 \epsilon} +  \left( \ln \frac{\sqrt{\pi}T e^{- \gamma_E/2}}{2 \nu}  -1 \right)
+  \frac 14 + \left( 1+\frac{L^2}{4 l^2}\right) \left(1-\frac{l_0}{2 l}
 \,{\rm ln\,} \frac{l_0+l}{l_0-l}  \right)  \right] \,,
\label{final-T}
\end{align}
where 
one can see that the pole is local, and agrees with that of the photon wave function renormalization.

\end{document}